\documentclass[aps,prr,twocolumn,superscriptaddress,longbibliography]{revtex4-2}

\usepackage{amsmath,amssymb,graphicx,bm}
\usepackage{physics}

\begin{document}

\title{Transient Information Partition in Coherent Exciton–Phonon–Photon Dynamics}

\author{Kunio Ishida}
\affiliation{School of Engineering and Center for Optical Research and Education, Utsunomiya University, Utsunomiya 321-8585, Japan}

\date{\today}

\begin{abstract}
We study transient information partition in a coherent exciton--phonon--photon system using subsystem-resolved quantum mutual information (QMI).
By employing a model with excitonic, phononic, and photonic degrees of freedom, we analyze the dynamics in the $J$-$\nu$ plane, where $J$ characterizes excitonic delocalization and $\nu$ denotes the exciton--phonon coupling strength.
By comparing time-averaged QMI maps with the absorbed photon number, we show that optical activity alone does not determine the character of the light-induced transient state.
The exciton-centered information partition identifies a broad crossover between polariton-like and polaron-like transient responses, depending on whether excitonic information is mainly shared with the photon or phonon subsystem.
In contrast, the phonon-centered partition reveals a sharper boundary-adjacent redistribution ridge near the boundary between the zero- and one-exciton ground-state sectors.
This ridge is absent from both the ground-state sector map and the photon-absorption map, indicating that it is neither a static sector boundary nor an enhancement of optical absorption.
A variational strength-function analysis connects the ridge to a region-II-like finite-energy polaronic excitation whose dominant spectral weight lies near the one-phonon energy, and thus the ridge represents a hidden transient correlation structure in which a limited amount of phonon-related information is preferentially shared with the photon subsystem before being predominantly allocated to exciton--phonon dressing.
These results show that QMI-based information partition provides a correlation-based framework for characterizing coherent light-induced transient states in which optical and material degrees of freedom jointly participate quantum mechanically.
\end{abstract}

\maketitle

\section{Introduction}

Light irradiation has long been used as a means of controlling material states, including coherent control of photochemical reactions and collective phenomena such as photoinduced phase transitions \cite{Koshihara2022PhysRep,DeLaTorre2021RMP}.
In many studies on such processes, the light field is treated as a classical external drive, and the material response is characterized in terms of excitation densities, lattice displacements, order parameters, or conventional correlation functions.
However, in the short-time coherent regime, the photon field and the material degrees of freedom can become quantum mechanically correlated during the transient evolution.
It is then necessary to characterize not only the quantum state of the material subsystem, but also the correlated light--matter state as a whole.
This viewpoint is also relevant to the broader idea of on-demand quantum materials \cite{Basov2017NatMater}, where light is used not only to probe a preexisting material state but also to create and control transient quantum states.

This problem naturally arises in systems where several correlation channels coexist.
In electron--phonon systems, exciton-polaron formation provides a standard quasiparticle picture in which an excitonic excitation is dressed by lattice displacement.
This picture is supported by Holstein-type models of local electron--phonon coupling and by recent first-principles studies showing that exciton--phonon interactions can produce substantial dressing, localization, and self-trapping effects \cite{Holstein1959AnnPhys,AntoniusLouie2022PRB,Dai2024PRL,Tao2021NatCommun}.
In molecular aggregates and related Frenkel-exciton systems, vibronic sidebands further show that excitonic delocalization and vibrational structure can occur on comparable energy scales \cite{EisfeldBriggs2006ChemPhys,SpanoYamagata2011JPCB,JumboNogales2022JPCL}.
On the other hand, in light--matter coupled systems, exciton-polariton formation describes coherent hybridization between matter excitations and photon modes.
Organic and molecular polaritonic systems have also been modeled using Dicke--Holstein-type descriptions, where vibrational dressing and light--matter coupling jointly affect the crossover between weak and strong coupling regimes \cite{Strashko2018PRL,Hertzog2019ChemSocRev}.

These quasiparticle concepts have provided useful reference languages for exciton--phonon and exciton--photon correlations.
In a coherent photoirradiated transient state, however, the relevant correlations are generated within the same many-body wave function.
In this case, exciton-polaron-like and exciton-polariton-like correlations can coexist and be redistributed in time, rather than forming mutually exclusive dynamical sectors.
Thus, assigning a single quasiparticle character to the transient state is not sufficient.
The central question is how the correlations generated by light irradiation are partitioned among excitonic, phononic, and photonic degrees of freedom.

This motivates a more general description in terms of correlation structure.
Quantum mutual information (QMI) provides a useful basis for such a description because it quantifies the total correlation shared by two selected subsystems and allows different correlation channels to be compared on the same footing.
In interacting many-body systems, QMI and related information-theoretic quantities have been used to characterize ground-state and many-body correlations \cite{DeChiara2018RPP,DeTomasi2017PRL}.
Multipartite information measures have also been used to analyze correlation structure, scrambling, and the spreading of quantum information in many-body dynamics \cite{Seshadri2018PRE,Banuls2017PRB,IyodaSagawa2018PRA}.
These studies show that information-theoretic quantities can be used to characterize many-body states without selecting particular microscopic correlation operators in advance.

Related developments have appeared in spin-boson and bosonic-environment models, where numerically controlled dynamical approaches have been used to analyze nonequilibrium dynamics and information-based diagnostics \cite{PineiroOrioli2017PRA,Suarez2024PRA,Goulko2025PRL,Parlato2025PRB}.
These studies are relevant to the present work because they treat bosonic degrees of freedom explicitly and analyze their nonequilibrium dynamics.
The present model differs from such settings in that the phonon and photon modes are not regarded as an environment to be integrated out.
Instead, they are treated as coherent quantum subsystems participating in the unitary transient dynamics.
QMI has also been applied to photoinduced electron--phonon dynamics, where it resolves the generation and transfer of correlations among electronic, phononic, and photonic degrees of freedom during coherent time evolution \cite{IshidaMatsueda2021JPSJ,IshidaMatsuedaKamada2022Faraday}.
This motivates using QMI not only as a measure of pairwise correlation, but as a way to analyze how optically generated correlations are partitioned among multiple physical degrees of freedom.

In this paper, we address this problem using a fully quantum exciton--phonon--photon model.
Since the exciton bandwidth $J$ and the exciton--phonon coupling $\nu$ determine the balance between excitonic delocalization and phonon dressing, we analyze the dynamics in the $J$-$\nu$ plane.
The ground-state sector map in this plane provides a structural reference for the dynamics, while the time-averaged QMI maps characterize the transient correlation structure generated after photon excitation.
By comparing these QMI maps with the absorbed photon number, we show that optical activity alone does not determine the character of the light-induced transient state.
We further introduce normalized QMI as an information partition, which measures how the information associated with a chosen subsystem is distributed among the other subsystems.
With the exciton subsystem as the reference, this analysis identifies a broad crossover between polariton-like and polaron-like transient responses, depending on whether excitonic information is mainly shared with the photon or phonon subsystem.
With the phonon subsystem as the reference, the same analysis reveals a sharper boundary-adjacent redistribution ridge near the I/II boundary on the region-I side.
This ridge is absent from both the ground-state sector map and the photon-absorption map, and therefore represents a hidden transient correlation structure rather than a static sector boundary or an enhancement of optical absorption.

The present results show that coherent light irradiation redistributes correlations not only within the material sector but also between material and photonic degrees of freedom.
The QMI-based information partition distinguishes optical activity from correlation redistribution and identifies transient structures that are difficult to infer from conventional observables alone.
Although the present calculation is performed for a minimal four-site system, the same correlation-based framework is expected to be applicable to a broad class of coherent many-body systems composed of multiple interacting subsystems.

\section{Model and Method}

We consider a quantum system composed of interacting excitonic, phononic, and photonic degrees of freedom. 
The excitonic sector is described by a set of two-level systems, which we refer to as pseudospins, 
representing the presence or absence of a local excitation. 
This formulation allows us to treat electronic excitations as excitons and their coupling to bosonic modes on equal quantum footing.

The total Hamiltonian is written as \cite{IshidaMatsuedaKamada2022Faraday}
\begin{equation}
\mathcal{H}
=
\mathcal{H}_{\mathrm{ex}}
+
\mathcal{H}_{\mathrm{phonon}}
+
\mathcal{H}_{\mathrm{photon}}
+
\mathcal{H}_{\mathrm{ex\text{-}phonon}}
+
\mathcal{H}_{\mathrm{ex\text{-}photon}},
\label{eq:Htotal}
\end{equation}
where each term describes either the internal dynamics of a subsystem or the coupling between different subsystems.

The excitonic (pseudospin) sector is given by
\begin{equation}
\mathcal{H}_{\mathrm{ex}}
=
\sum_i
\varepsilon \,
\sigma_i^{+}\sigma_i^{-}
+
J
\sum_i
\vec{\sigma}_i \cdot \vec{\sigma}_{i+1},
\label{eq:Hex}
\end{equation}
where $\varepsilon$ is the local excitation energy and 
$\sigma_i^{\pm}=(\sigma_i^x \pm i\sigma_i^y)/2$ denotes the Pauli operator acting on the two-level system at site $i$. 
The parameter $J$ is the nearest-neighbor excitonic transfer parameter in the pseudospin representation and controls the delocalization of excitonic configurations.

The phonon degrees of freedom are modeled as quantized bosonic modes,
\begin{equation}
\mathcal{H}_{\mathrm{phonon}}
=
\sum_i
\omega \,
a_i^{\dagger} a_i ,
\label{eq:Hphonon}
\end{equation}
where $a_i$ ($a_i^\dagger$) annihilates (creates) a phonon at site $i$ with frequency $\omega$.

Similarly, the photon sector is described by
\begin{equation}
\mathcal{H}_{\mathrm{photon}}
=
\Omega \,
c^{\dagger} c ,
\label{eq:Hphoton}
\end{equation}
with $c$ and $c^\dagger$ denoting photon operators.

\begin{figure}
\scalebox{0.27}{\includegraphics*{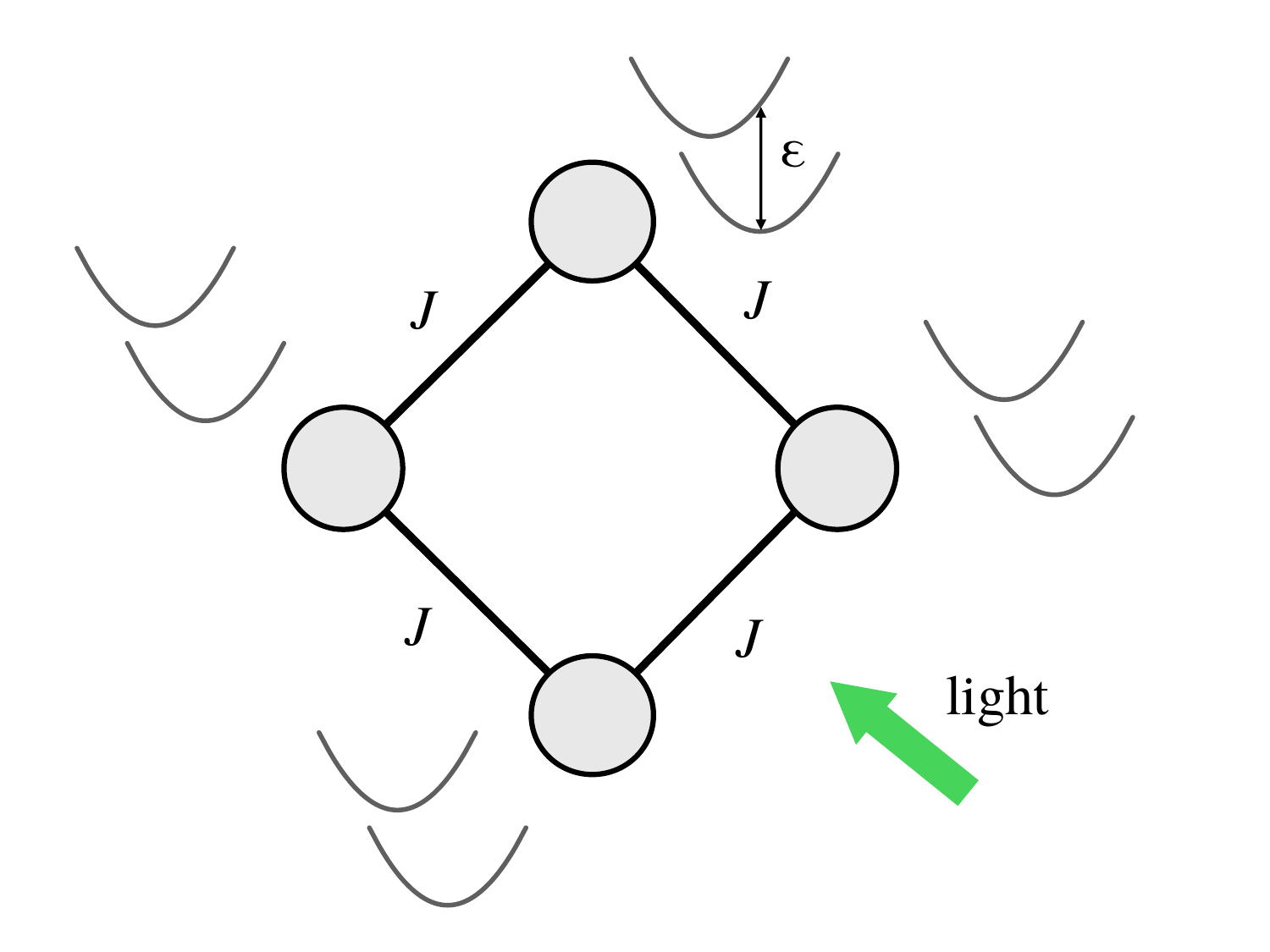}}
\caption{A schematic view of the model.}
\label{fig:schem}
\end{figure}

The exciton--phonon coupling is modeled by a local Holstein-type interaction, which represents a minimal description of excitation-induced lattice displacement \cite{Holstein1959AnnPhys}. 
This form is consistent with the broader electron--phonon coupling framework used in molecular and solid-state systems \cite{Giustino2017RMP} and with vibronic descriptions of Frenkel-exciton aggregates \cite{SpanoYamagata2011JPCB}, and thus the corresponding Hamiltonian is given by
\begin{equation}
\mathcal{H}_{\mathrm{ex\text{-}phonon}}
=
\sum_i
\nu
\sigma_i^{+}\sigma_i^{-}
(a_i^{\dagger}+a_i),
\label{eq:Hexphonon}
\end{equation}
and
\begin{equation}
\mathcal{H}_{\mathrm{ex\text{-}photon}}
=
\sum_i
\mu
\sigma_i^{x}
(c^{\dagger}+c),
\label{eq:Hexphoton}
\end{equation}
which describe excitation-induced lattice distortions and light--matter coupling, respectively. 
This form encompasses a wide class of exciton--phonon and exciton--photon interactions in correlated optical media.
The model is schematically described in Fig.\ \ref{fig:schem}.

Throughout this work, the phonon and photon modes are treated as fully quantum mechanical degrees of freedom. 
Importantly, they are not regarded as an environment inducing dissipation, but as coherent subsystems 
participating in the unitary dynamics of the total system.

The time evolution is therefore governed by the Schr\"odinger equation
\begin{equation}
|\Psi(t)\rangle
=
e^{-i\mathcal{H}t}
|\Psi(0)\rangle ,
\label{eq:time}
\end{equation}
where $|\Psi(0)\rangle$ is an initial pure state of the composite system.

This formulation naturally defines a partition of the total Hilbert space into excitonic (pseudospin), 
phononic, and photonic subsystems. In the following sections, this partition will be used to analyze 
the information structure of transient quantum states, without assuming proximity to equilibrium 
or adiabatic continuity to ground-state properties.

Equation~(\ref{eq:time}) is numerically solved by the Krylov subspace method \cite{ParkLight1986JCP,Saad1992SJNA}.
We take a four-site system with periodic boundary conditions in which the photon occupation is truncated at $m_{\max}=31$, and each local phonon occupation is truncated at $n_{\max}=15$.
The Krylov subspace method with the Lanczos algorithm \cite{ParkLight1986JCP} is applied to obtain an orthogonal basis set at each time, and the rank of the Krylov subspace is fixed to $32$. The time step $\Delta t$ is $0.028$, which we confirm to be sufficient to obtain numerical convergence for the present model \cite{supp}.

In this study, the local phonon frequency $\omega$ is taken to be the unit of energy  and $\hbar=1$.
We fix $\omega=1$, $\varepsilon=\Omega=13.5$, and scan the $J$-$\nu$ plane.
The values of $\varepsilon$ and $\Omega$ correspond to an optical excitation whose energy is one order of magnitude larger than a characteristic optical-phonon or intramolecular vibrational energy.
The exciton hopping $J$ is varied over $J=O(1)$, corresponding to a regime in which excitonic delocalization competes with vibronic structure, as commonly discussed in Frenkel-exciton molecular aggregates \cite{EisfeldBriggs2006ChemPhys,SpanoYamagata2011JPCB,JumboNogales2022JPCL}.
The exciton--phonon coupling $\nu$ is also varied over $\nu=O(1)$, where phonon dressing can lead to exciton-polaron formation and self-trapping behavior \cite{Holstein1959AnnPhys,AntoniusLouie2022PRB,Dai2024PRL,Tao2021NatCommun}.
The photon--matter coupling $\mu=0.5$ is treated as an effective coupling parameter that depends on the optical mode and excitation condition.
In molecular polaritonic settings, weak and strong light--matter coupling are controlled not only by microscopic transition dipoles but also by mode confinement, collective enhancement, and linewidths \cite{Strashko2018PRL,Hertzog2019ChemSocRev}.
Here $\mu=0.5$ is chosen to produce appreciable coherent photon-induced dynamics without driving the system into a saturated absorption regime, so that photon absorption and correlation redistribution can be compared within the same parameter window.

The initial state of the entire system is a tensor product of the ground state of the electron--phonon system and a coherent state of photons.
The average number of photons is 14 at $t=0$, and the simulation is performed from $t=0$ to $t=T=7$.
All parameter-space maps were evaluated on a uniform grid with spacings $\Delta J=\Delta \nu =0.2$, and the color maps were obtained by interpolating the calculated grid values for visualization.

\section{Calculated Results}

\subsection{Ground-state classification}
\label{sec:gsmap}

\begin{figure}[htbp]
\scalebox{0.24}{\includegraphics*{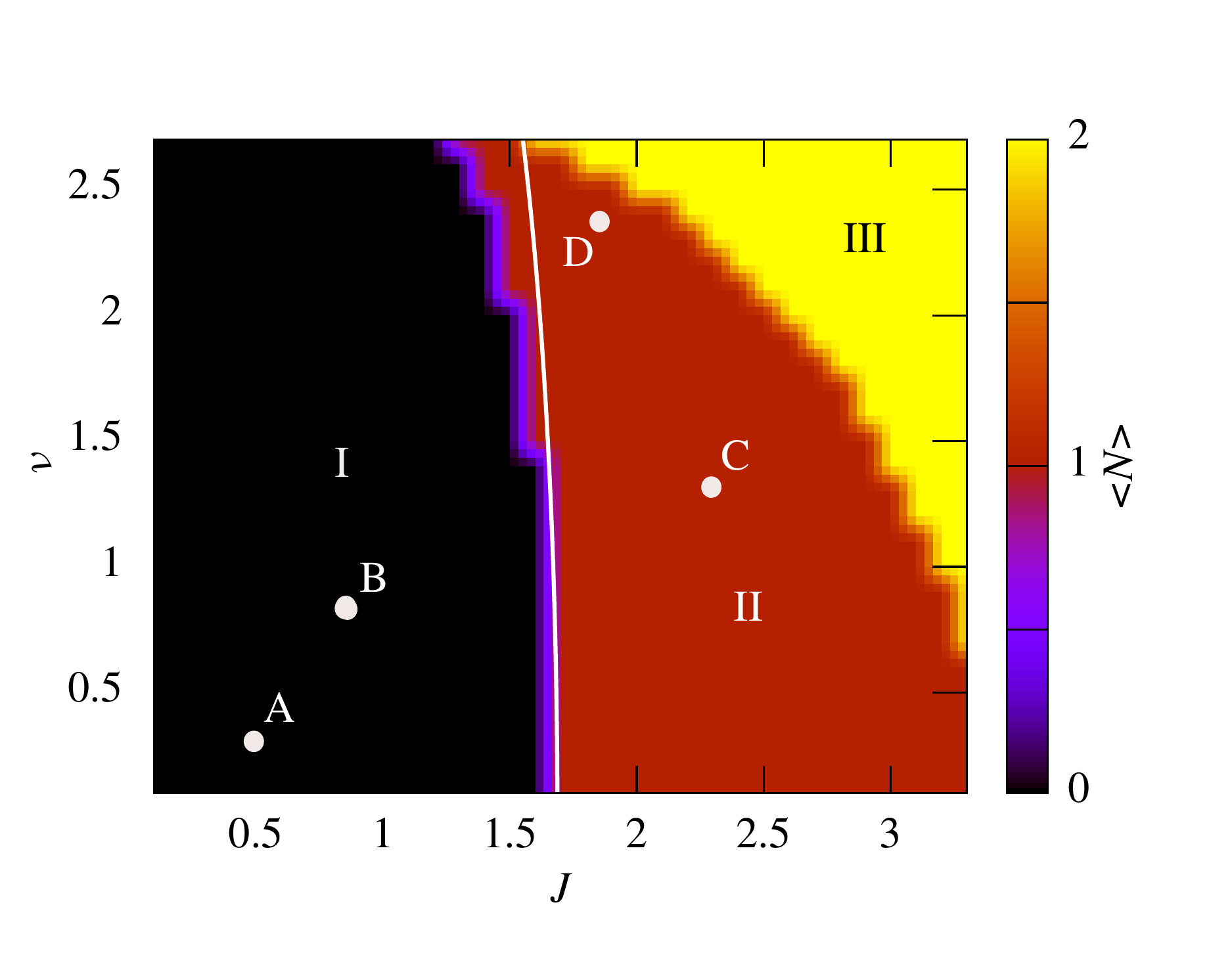}}
\caption{Ground-state classification map. The regions I, II, and III correspond predominantly to $N=0$, $1$, and $2$, respectively. The symbols A-D in the map indicate the four parameter points used for the time traces in Fig.~\ref{fig:partition_time}. The white line is calculated by the minimal variational polaron ansatz.}
\label{fig:gs_map}
\end{figure}

Before studying the dynamical properties, we first discuss the ground-state structure in the absence of photons and focus on the competition between $J$ and $\nu$.
We first note that the number of excitons $N = \sum_{i}\sigma^+_i \sigma^-_i$ is conserved since $[{\cal H}_{\mathrm{ex}} + {\cal H}_{\mathrm{phonon}} + {\cal H}_{\mathrm{ex\text{-}phonon}}, N] = 0$.
Figure~\ref{fig:gs_map} shows the excitation-number map in the $(J,\nu)$ plane, where the color denotes the value of $\langle N \rangle = \bra{\psi} N \ket{\psi}$ with the ground-state wavefunction $\ket{\psi}$.
We identify three regions, I, II, and III, according to the dominant excitation number, corresponding to $N=0$, 1, and 2, respectively.

For small $J$ and $\nu$, the excitation cost $\varepsilon$ dominates over the kinetic gain, and the ground state remains in the $\langle N \rangle =0$ sector.
Therefore, no excitons or phonons are present in region I and $\ket{\Psi_{\rm I}} = \ket{- - - - 0 0 0 0}$.
As a result, the ground-state energy in region I is given by $4J$.

As the value of $J$ increases, exciton creation becomes more preferable and thus a single exciton appears in region II.
Furthermore, when an exciton is created, the energy gain by the exciton-phonon interaction lowers the ground-state energy, and thus the boundary between region I and II shifts to the left as $\nu$ increases.
This feature is understood by a variational wavefunction in region II, which is given by
\begin{widetext}
\begin{equation}
\ket{\Psi_{\rm II}} = \frac{1}{2} \left ( \ket{+ - - - \alpha 0 0 0 }- \ket{- + - - 0 \alpha 0 0 } + \ket{- - +  - 0 0 \alpha 0} - \ket{- - - + 0 0 0 \alpha} \right ),
\label{var}
\end{equation}
\end{widetext}
where $\alpha$ denotes the parameter for a phonon coherent state at corresponding sites.
$\ket{\Psi_{\rm II}}$ shows that a single exciton dressed by phonons is present with translational symmetry ($k=\pi$), and thus we refer to this state as a minimal variational polaron ansatz.

The variational ground state in region II is given by minimizing the energy $E_{\rm II} = \langle\Psi_{\rm II}| {\cal H} | \Psi_{\rm II}\rangle $ given by 
\begin{equation}
E_{\rm II} = -4Je^{-\alpha^2} + \varepsilon + 2 \nu \alpha + \omega \alpha^2.
\label{e2}
\end{equation}
Comparing Eq.\ (\ref{e2}) with the ground-state energy in region I, I/II boundary is given by
\begin{equation}
4J = -4Je^{-\alpha_0^2} + \varepsilon + 2 \nu \alpha_0 + \omega \alpha_0^2,
\label{bnd}
\end{equation}
where $\alpha_0$ minimizes $E_{\rm II}$ and satisfies
\begin{equation}
\alpha_0 (4Je^{-\alpha_0^2} + \omega) + \nu = 0.
\label{e2min}
\end{equation}
For $\nu \sim 0$, $\alpha_0$ is also small and Eq.\ (\ref{e2min}) gives
\begin{equation}
\alpha_0 \sim -\nu/(4J+\omega).
\label{alpha0}
\end{equation}
Substituting Eq.\ (\ref{alpha0}) into Eq.\ (\ref{bnd}), we obtain 
\begin{equation}
\nu^2 \sim (\varepsilon - 8J)(\omega + 4J),
\label{approxbd}
\end{equation}
to the second order of $\nu$, which shows the boundary around $(J, \nu)=(\varepsilon/8, 0)$.

Numerically calculated curve of the boundary is shown by the line in Fig.\ \ref{fig:gs_map}, which agrees well with the boundary inferred from $\langle N \rangle$.
These results show that the lattice distortion at the boundary is suppressed as $J$ increases, which means that the energy gain owing to the exciton-phonon interaction decreases.
Hence, the boundary line shifts towards smaller $J$ in Fig.\ \ref{fig:gs_map}.
Within the minimal variational polaron ansatz Eq. (\ref{var}),
the I/II boundary can be interpreted as an energetic
balance between the kinetic gain and the phonon-induced dressing.
In particular, the boundary approximately follows Eq.\ (\ref{approxbd}) in the weak-dressing limit. 

We note that the local energetic mechanism underlying the I/II boundary is not specific to the four-site cluster and is expected to persist, at least qualitatively, in larger finite systems.
For $2M$-site systems, the relevant ground-state sectors are characterized by the number of excitons $k$. 
In the parameter range considered here, the low-energy sequence runs through $k = 0$, $1$, $2$, ..., $M$, where the boundaries at the extremal sectors are governed solely by the competition between the on-site excitation energy $\varepsilon$ and the transfer energy $J$ in the absence of exciton–phonon coupling ($\nu = 0$).

In particular, the boundary between $k=0$ and $k=1$ is determined by the lower edge of the single-exciton band.
This reflects the local nature of the energy competition: the boundary is controlled by the balance between the excitation cost $\varepsilon$ and the lattice-relaxation gain $\nu^2$, while the contribution by exciton transfer energy enters only through the incremental energy change associated with adding one excitation.
This argument suggests that the mechanism determining the I/II boundary is not specific to the four-site cluster.

Although the present study focuses on a four-site system ($M=2$), this sector-based classification provides a natural organizing principle that extends beyond small system sizes.
As $M$ increases, the sequence of level crossings between neighboring excitation sectors is expected to become denser, so that the excitation density varies more smoothly with the model parameters.
In the following, we employ this finite-size sector structure as a reference framework for analyzing the dynamical redistribution of quantum correlations.

We also stress that this classification is purely static.
It identifies the energetically favored excitation sector, though it does not determine how correlations are redistributed among excitonic, phononic, and photonic degrees of freedom after light irradiation.
To uncover such dynamical aspects, we now turn to the transient information structure.

\subsection{Transient correlation maps and exciton-centered information partition}
\label{sec:correlationmaps}

We next discuss the transient dynamics of the system under light irradiation.
Our purpose is to compare two structures within the same fully quantum model: the ground-state sector map and the transient correlation structure generated after photon excitation.
The four-site system provides a minimal setting in which this comparison can be made while retaining excitonic, phononic, and photonic degrees of freedom as quantum subsystems.

For a time-dependent quantity $X(t)$, we define
\begin{equation}
\overline{X}
=
\frac{1}{T}
\int_0^T X(t)\,dt ,
\end{equation}
where $T$ is the time window used for the averaging. 
We denote each subsystem by $e$, $pn$, and $pt$ for excitons, phonons, and photons, respectively.
The time-averaged pairwise QMI and subsystem entropy are denoted by
$I_{a,b}$ and $S_a$, respectively, with $a,b\in\{e,pn,pt\}$.

\begin{figure}
\scalebox{0.24}{\includegraphics*{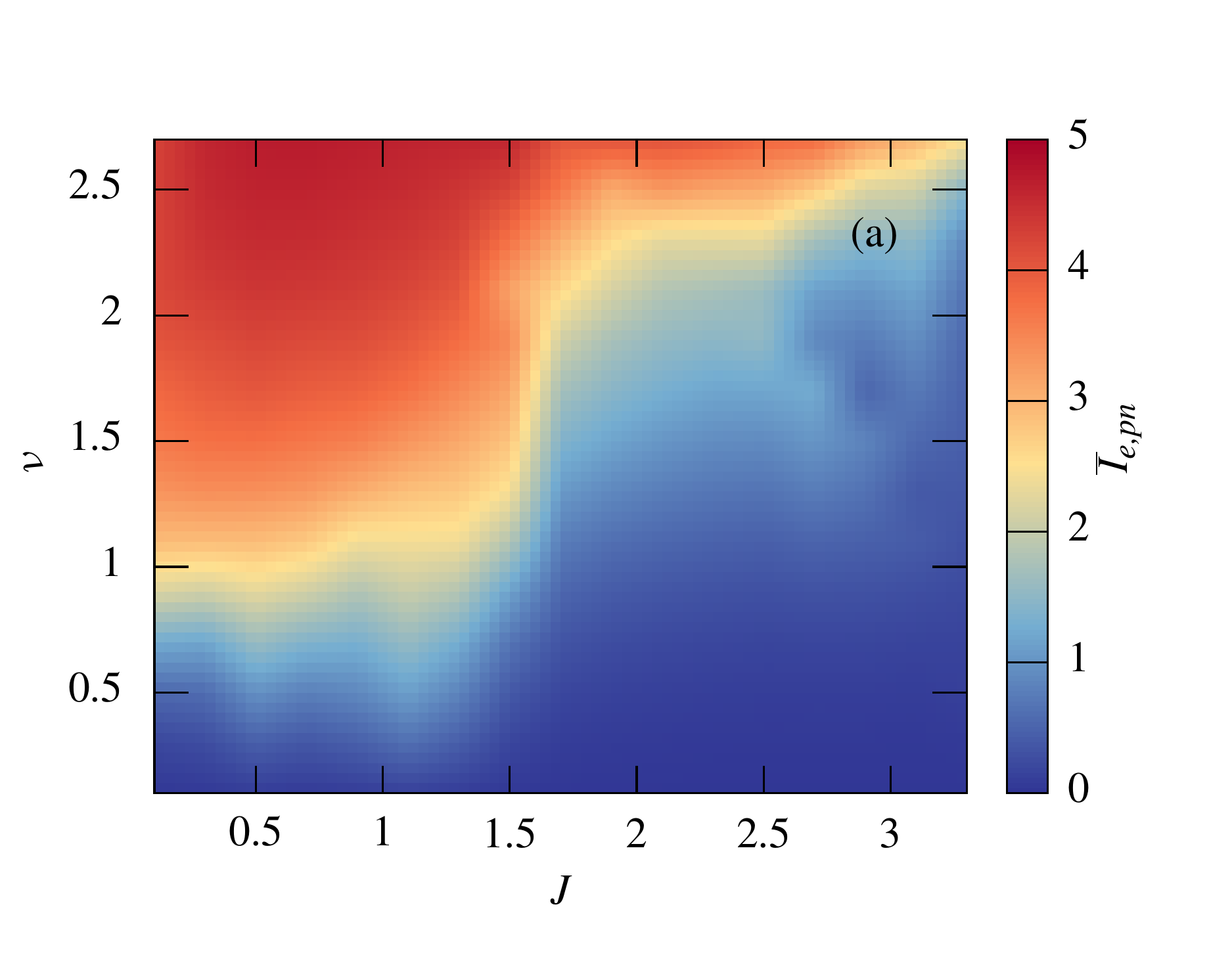}}
\scalebox{0.24}{\includegraphics*{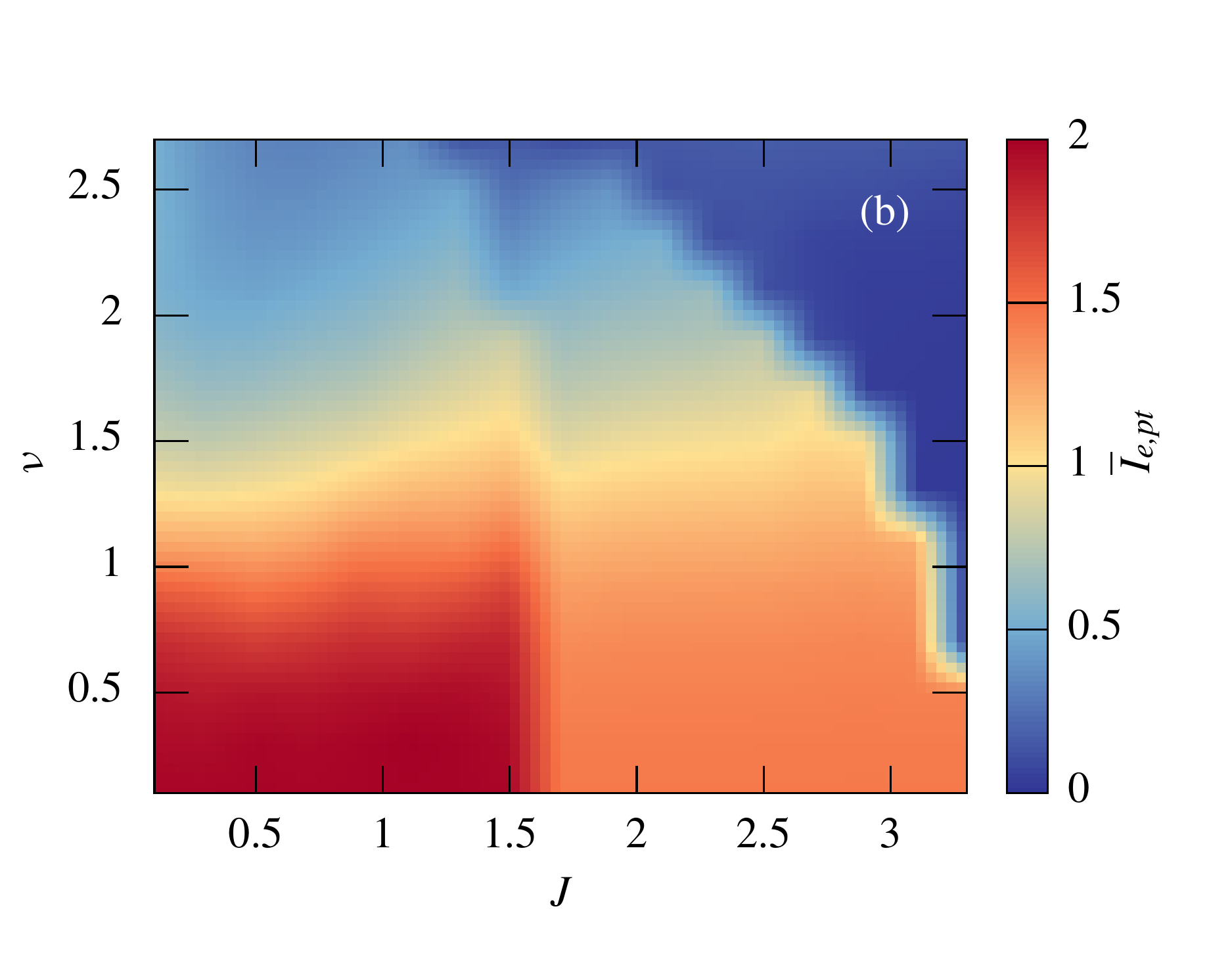}}
\scalebox{0.24}{\includegraphics*{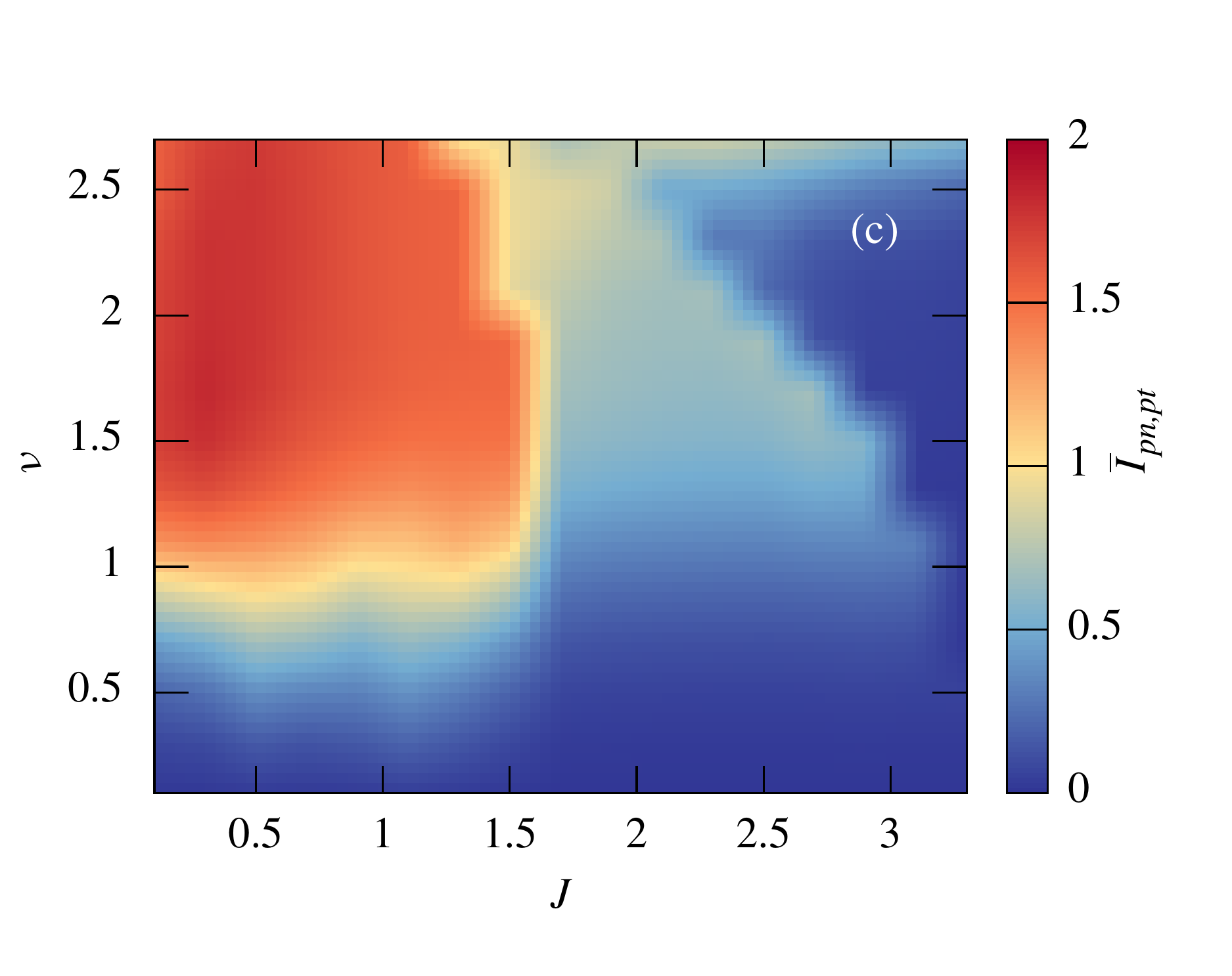}}
\scalebox{0.24}{\includegraphics*{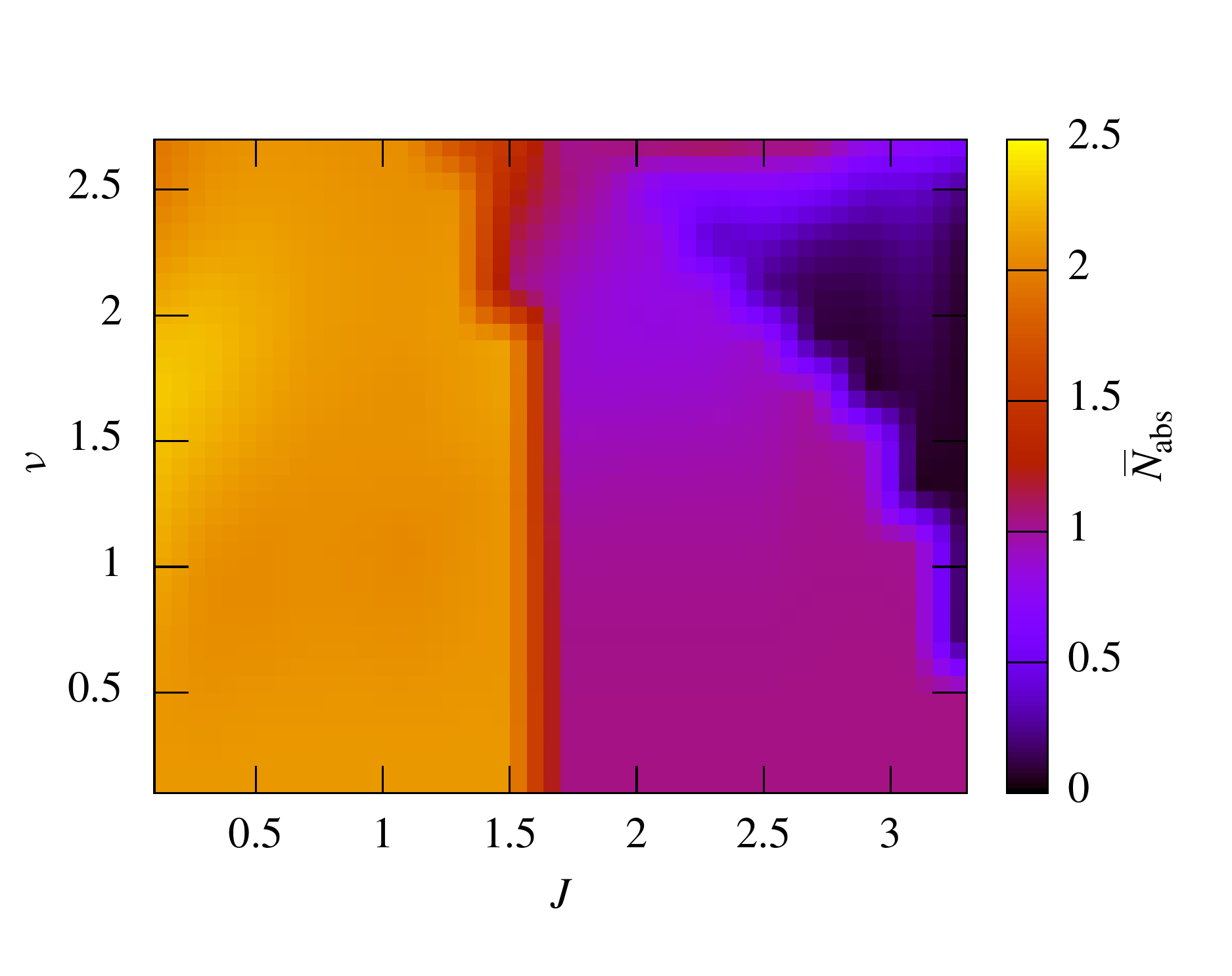}}
\caption{Time-averaged QMI and absorbed-photon maps in the $J$-$\nu$ plane.
(a) $\overline{I}_{e,pn}$, (b) $\overline{I}_{e,pt}$, and
(c) $\overline{I}_{pn,pt}$.
(d) Time-averaged absorbed photon number $\overline{N}_{\rm abs}$.}
\label{fig:qmi_maps}
\end{figure}
Figures~\ref{fig:qmi_maps}(a)-(c) show the time-averaged QMI maps
$\overline{I}_{e,pn}$, $\overline{I}_{e,pt}$, and $\overline{I}_{pn,pt}$
in the $J$-$\nu$ plane. 
Although these maps are computed from the time evolution after photon excitation, the initial state at each parameter point is the corresponding ground state.
The transient QMI maps show broad region-dependent features associated with the ground-state sectors, although they are not identical to the ground-state map.
They therefore provide a different view of the transient state.
More specifically, the exciton--phonon QMI $\overline{I}_{e,pn}$ is enhanced toward the phonon-dressed side of the diagram, whereas the exciton--photon QMI $\overline{I}_{e,pt}$ is more visible in the photon-accessible region.
The phonon--photon QMI $\overline{I}_{pn,pt}$ also develops a distinct parameter dependence, although the two subsystems are not directly coupled in the Hamiltonian.
Thus, the transient state contains a correlation structure that is not resolved by the ground-state classification alone.

The three QMI components give different views of the same transient dynamics.
Region III remains relatively insensitive to the present photon-excitation scheme and therefore acts as a nearly inactive sector in the QMI maps.
Region I shows a comparatively simple change from a photon-dominated (polaritonic) response at small $\nu$ to a more phonon-involved (polaronic) response as $\nu$ increases.

Region II is useful for separating the ground-state sector character from the transient correlation structure.
Although the ground-state analysis identifies this region as a polaron-like sector, the transient QMI maps do not reduce to a uniformly phonon-dominated response.
Photon-related correlations are generally weaker than in the photon-active part of region I, consistent with the reduced photonic activity.
At the same time, the phonon-related QMI does not immediately become dominant throughout region II; its onset is shifted to larger $\nu$ compared with region I.
Thus, the static polaronic character of region II constrains the transient dynamics.
However, it does not by itself determine how the photon-induced correlations are partitioned among the exciton, phonon, and photon subsystems.
This motivates an information-partition analysis of how the excitonic information is shared between the phonon and photon channels.

To connect the QMI maps with a conventional photonic observable, we also examine the time-averaged absorbed photon number, denoted by $\overline{N}_{\rm abs}$, as shown in Fig.~\ref{fig:qmi_maps}(d).
Here $\overline{N}_{\rm abs}$ is the reduction of the photon occupation from its initial value during the finite observation window defined by
\begin{equation}
\overline{N}_{\rm abs}
=
\overline{N}_{pt}(0)-\overline{N}_{pt} .
\end{equation}
This quantity characterizes the conventional optical response to the excitation protocol and provides a measure of photonic activity.
The absorption map largely follows the ground-state sector structure: the photon channel is more active in region I and is suppressed in the more polaronic region II and in region III.
Thus, the absorption map indicates where the photon channel participates in the dynamics, whereas QMI specifies how the correlations generated through this optical coupling are shared among the exciton, phonon, and photon subsystems.

\begin{figure}
\scalebox{0.24}{\includegraphics*{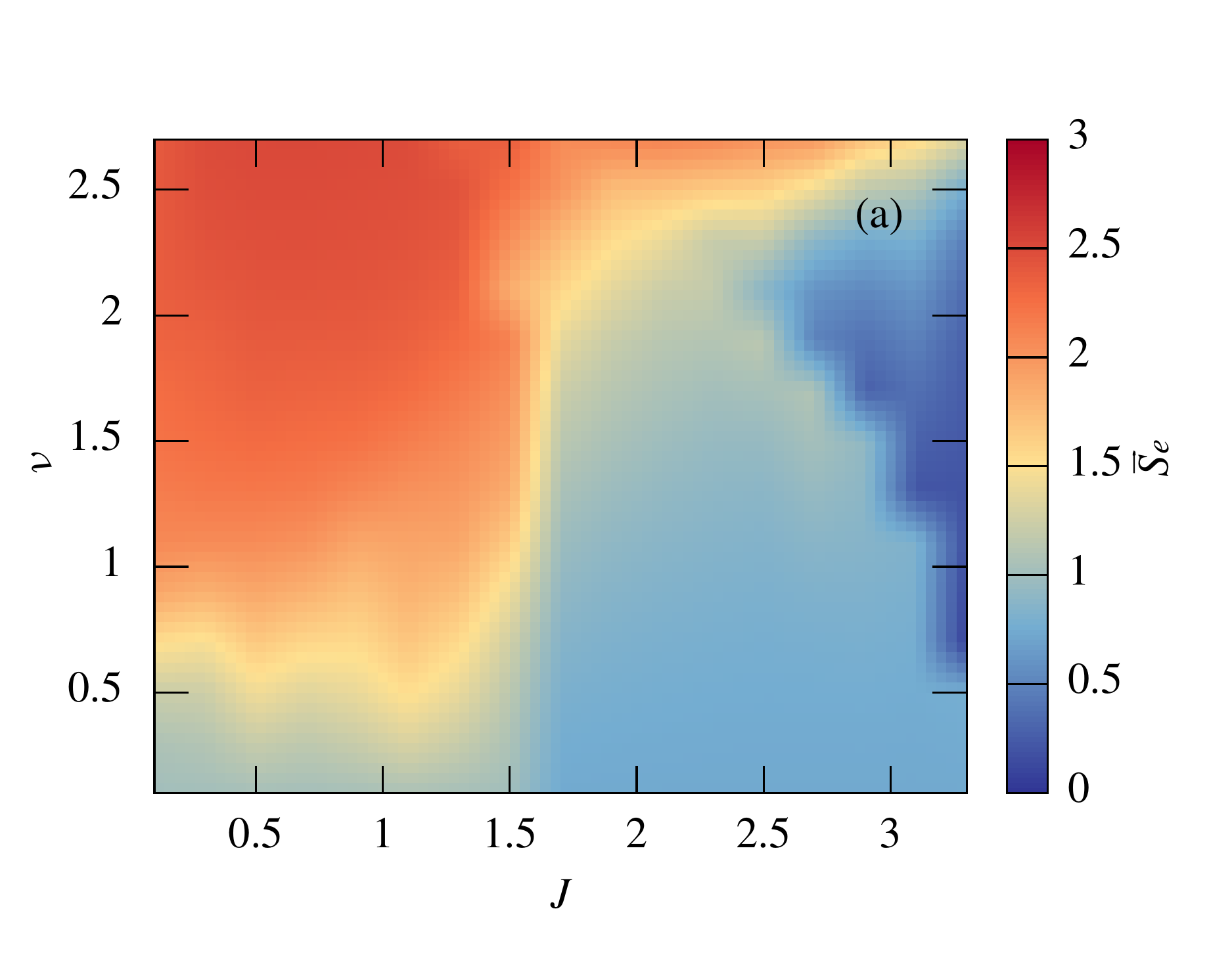}}
\scalebox{0.24}{\includegraphics*{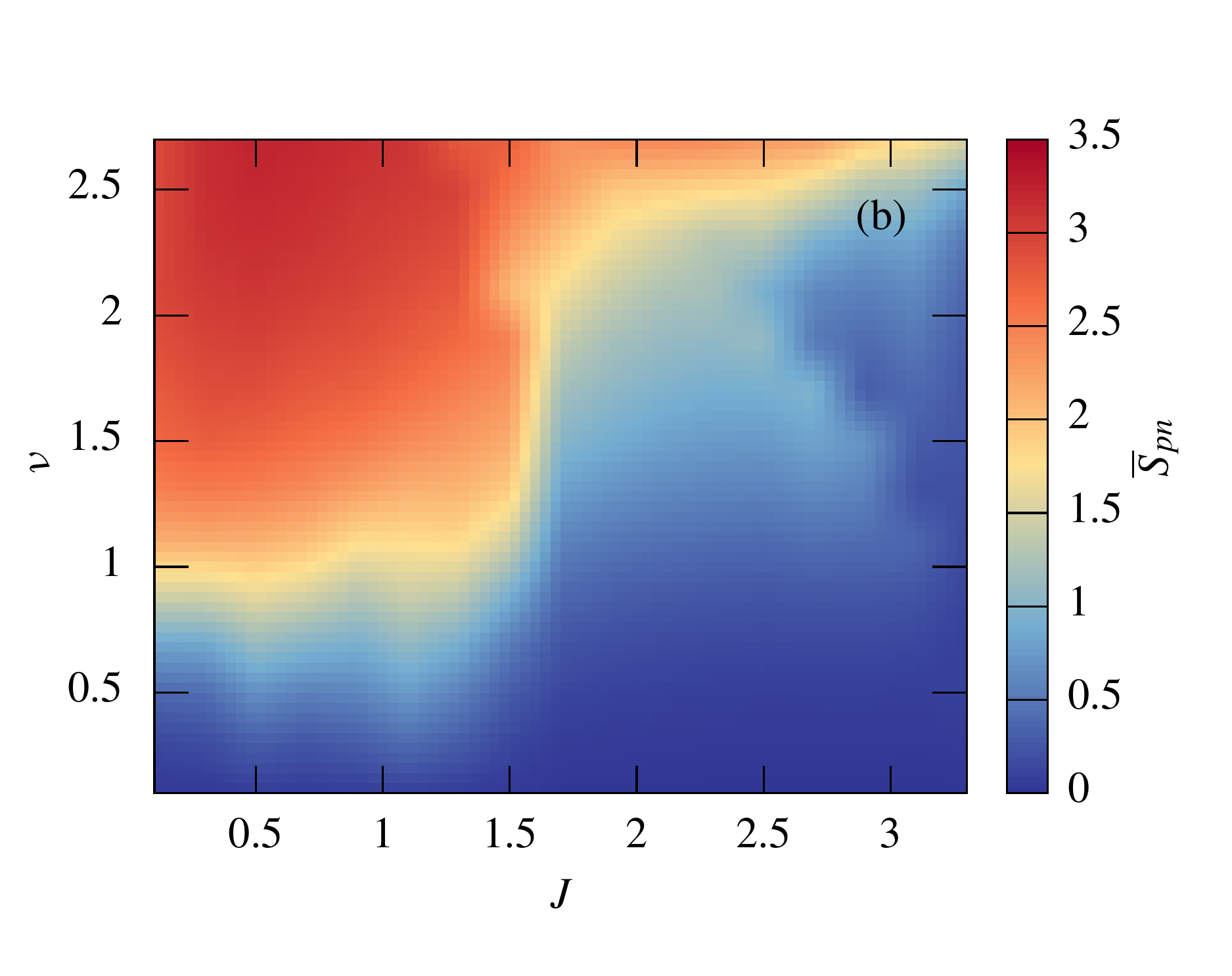}}
\scalebox{0.24}{\includegraphics*{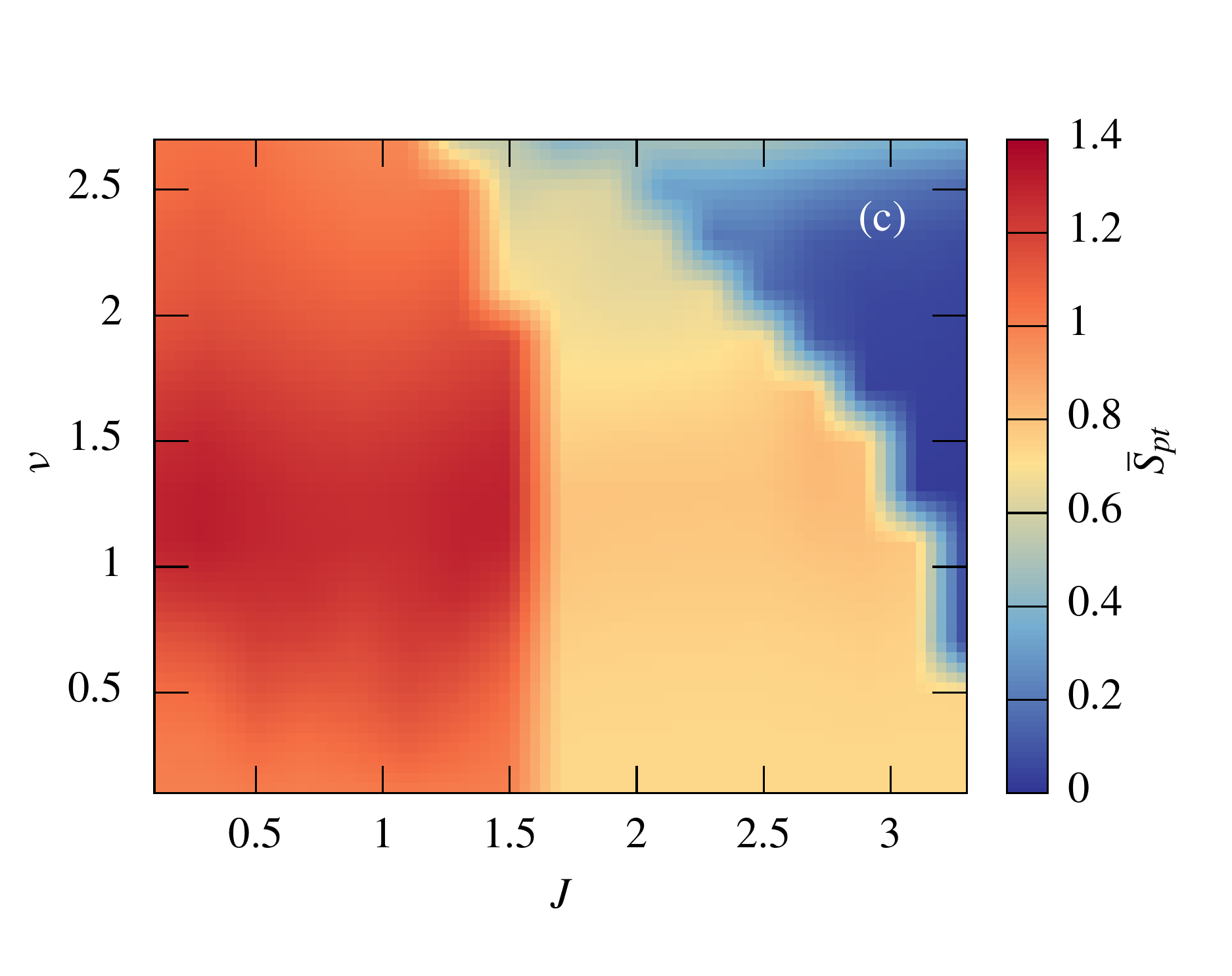}}
\caption{Time-averaged subsystem entropies in the $J$-$\nu$ plane.
(a) $\overline{S}_e$, (b) $\overline{S}_{pn}$, and (c) $\overline{S}_{pt}$.}
\label{fig:entropy_partition}
\end{figure}
The entanglement entropies of the subsystems, $\overline{S}_e$,
$\overline{S}_{pn}$, and $\overline{S}_{pt}$, are shown in
Fig.~\ref{fig:entropy_partition}. These maps provide a complementary
information-theoretic view of subsystem participation. For example, the
photon entropy $\overline{S}_{pt}$ follows the same broad tendency as the
photon absorption, indicating that both quantities capture the overall
participation of the photon subsystem. Thus, $\overline{S}_{pt}$, together
with $\overline{N}_{\rm abs}$, provides a useful indicator of
photon-channel accessibility: it is reduced on the more polaronic side of
the diagram and remains appreciable on the region-I side of the I/II
boundary. In particular, the exciton and phonon entropies remain finite over
broad parts of region II, while the pairwise QMI components vary
substantially.

While the entanglement entropy characterizes the information content of a
single subsystem, QMI quantifies the information shared between two selected
subsystems \cite{DeChiara2018RPP}. For a pure tripartite state composed of subsystems $A$, $B$,
and $C$, the QMI satisfies
\begin{equation}
I(A:B)+I(B:C)=2S(B).
\label{eq:qmi_equality}
\end{equation}
This identity allows $2S(B)$ to be interpreted as an information budget
of subsystem $B$, partitioned between its correlations with $A$ and
with $C$ \cite{Yang2023PRA}. Thus, although $\overline{S}_{pt}$ indicates the overall
participation of the photon subsystem, it does not specify whether the
photon-induced correlations are shared mainly with excitons or phonons.
The pairwise QMI resolves this partition.

Using Eq.~(\ref{eq:qmi_equality}) with $B=e$, we define the instantaneous
exciton-centered information partitions
\begin{equation}
\eta_{pn|e}(t)=\frac{I(e:pn;t)}{2S(e;t)},
\qquad
\eta_{pt|e}(t)=\frac{I(e:pt;t)}{2S(e;t)} ,
\label{eq:eta_ex_time}
\end{equation}
which satisfy $\eta_{pn|e}(t)+\eta_{pt|e}(t)=1$.
These quantities measure the instantaneous fractions of excitonic
information shared with phonons and photons, respectively.
We note that we evaluate $\eta_{pn|e}$ and $\eta_{pt|e}$ only for finite values of $S(e;t)$ in the numerical time traces.

\begin{figure}
\scalebox{0.7}{\includegraphics*{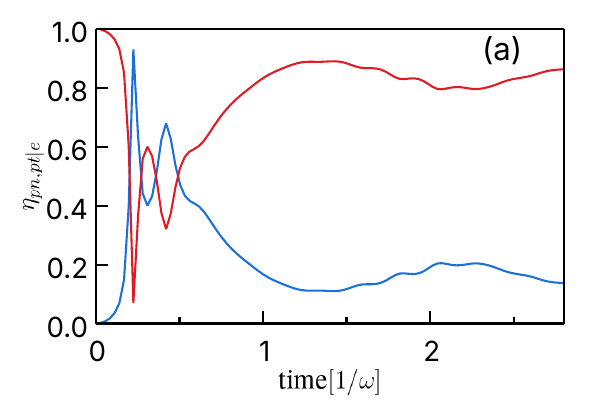}}
\scalebox{0.7}{\includegraphics*{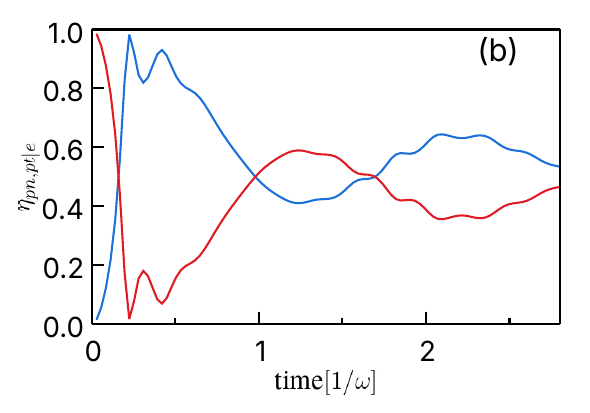}}
\scalebox{0.7}{\includegraphics*{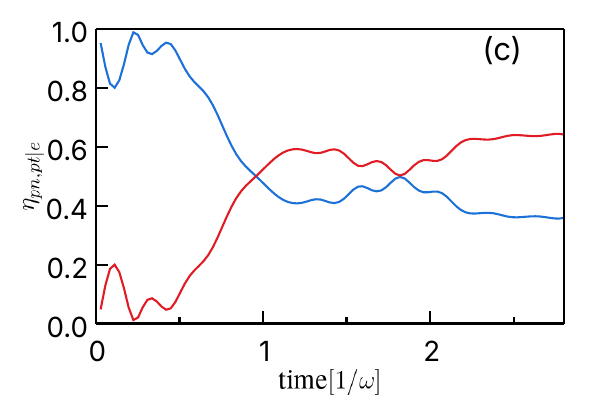}}
\scalebox{0.7}{\includegraphics*{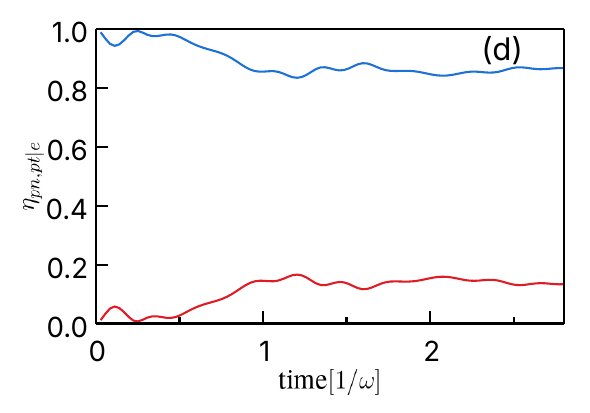}}
\caption{Time traces of the exciton-centered information partitions
$\eta_{pn|e}(t)$ (red line) and $\eta_{pt|e}(t)$ (blue line).
The four panels correspond to
(a) $(J,\nu)=(0.5,0.3)$,
(b) $(0.9,0.9)$,
(c) $(2.3,1.3)$, and
(d) $(1.9,2.3)$.}
\label{fig:partition_time}
\end{figure}
Figure~\ref{fig:partition_time} shows representative time traces of
$\eta_{pn|e}(t)$ and $\eta_{pt|e}(t)$ at four parameter points marked in Fig.~\ref{fig:gs_map}.
At $(J,\nu)=(0.5,0.3)$, the information partition changes rapidly during the early stage of the photon-induced dynamics, and $\eta_{pt|e}(t)$ remains dominant afterwards.
This point therefore represents a photon-dominated response in region I, where the excitonic information is mainly shared with the photon subsystem during the coherent evolution.
At $(J,\nu)=(0.9,0.9)$, which lies near the polariton--polaron-like crossover on the region-I side, $\eta_{pn|e}(t)$ and $\eta_{pt|e}(t)$ oscillate with comparable amplitudes, so that the dominant channel alternates in time.
This behavior shows that the crossover is not a static separation between two responses, but a regime of coherent competition between the phonon and photon channels.
The point $(J,\nu)=(2.3,1.3)$ lies in region II and illustrates that a static polaron-like sector does not immediately lead to a uniformly phonon-dominated transient response, i.e., a substantial photon-side contribution remains in the transient information partition.
At $(J,\nu)=(1.9,2.3)$, $\eta_{pn|e}(t)$ dominates, giving a clear phonon-dressed, polaron-like transient response.
These traces show that the time-averaged maps summarize genuine transient redistribution of excitonic information rather than simply reproducing the static ground-state classification.

\begin{figure}
\scalebox{0.24}{\includegraphics*{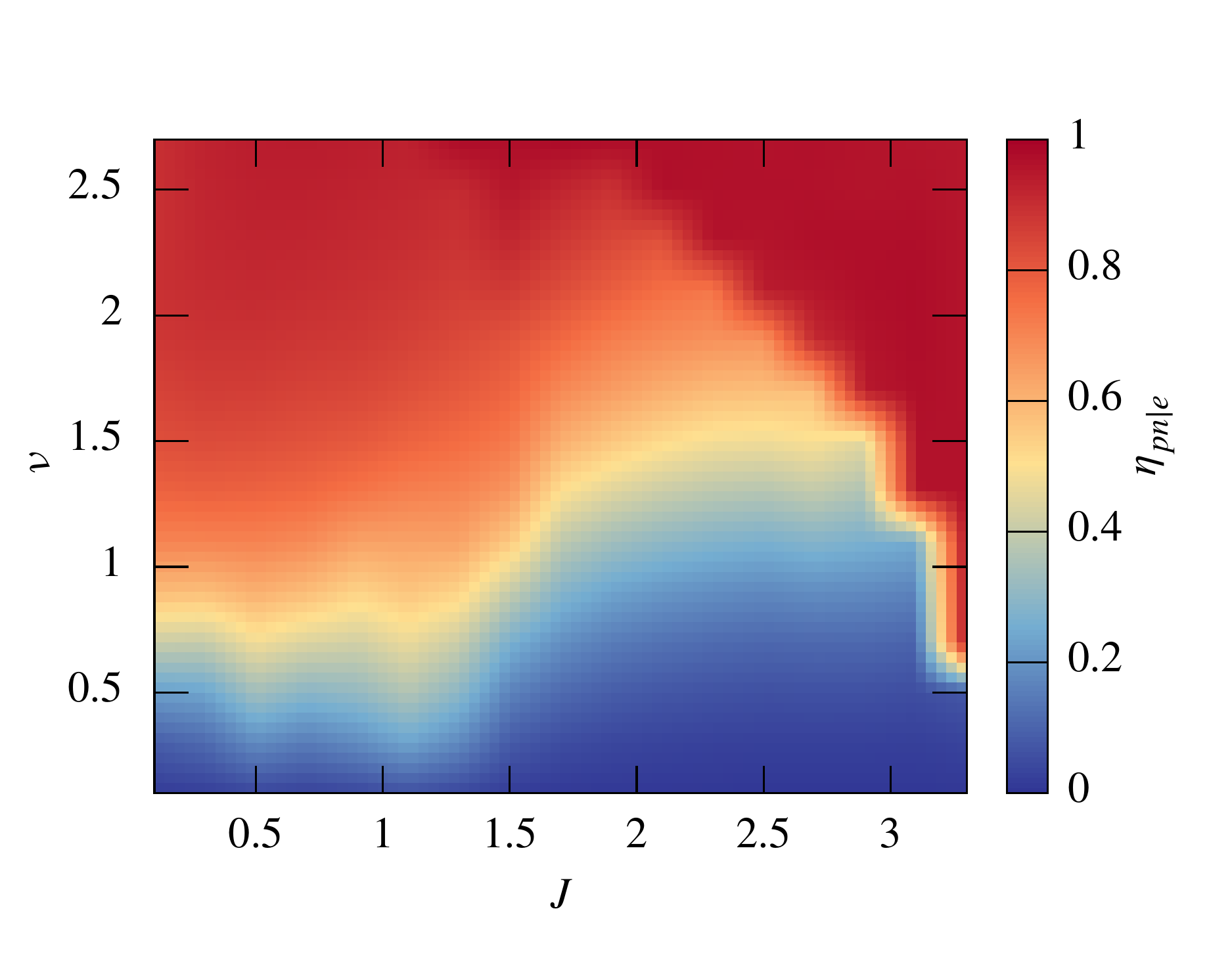}}
\caption{Exciton-centered information partition in the $J$-$\nu$ plane.
The color represents $\eta_{pn|e}$.
Small and large values correspond to photon-dominated and phonon-dominated transient responses, respectively.}
\label{fig:eta_ne}
\end{figure}

Since the same identity holds after applying the time average to each term,
we define the time-averaged exciton-centered information partitions as
\begin{equation}
\eta_{pn|e}
=
\frac{\overline{I}_{e,pn}}{2\overline{S}_e},
\qquad
\eta_{pt|e}
=
\frac{\overline{I}_{e,pt}}{2\overline{S}_e}.
\label{eq:eta_ex}
\end{equation}
The map of $\eta_{pn|e}$, shown in Fig.~\ref{fig:eta_ne},
reveals a broad crossover from photon-dominated to phonon-dominated
information sharing.
In the small-$\nu$ region, $\eta_{pt|e}$ is large and the transient state
has a polariton-like character in the sense that excitonic information is
mainly shared with the photon subsystem.
As $\nu$ increases, $\eta_{pn|e}$ grows and the excitonic information is
increasingly shared with the phonon subsystem, corresponding to a
phonon-dressed, polaron-like transient response.
The normalized QMI therefore provides a quantitative way to compare
polariton-like and polaron-like correlations on the same information scale.

This crossover should be distinguished from the static I/II boundary.
It extends continuously from region I into region II and therefore
represents a transient information-partition crossover rather than a
ground-state sector boundary.
In region I, where the ground state is essentially undressed, this crossover
can be interpreted as the gradual growth of phonon-dressed correlations from
a photon-accessible transient state.
In region II, the same crossover occurs on top of a static polaron-like
sector, and the onset of the phonon-dominated transient response is shifted
to larger $\nu$.

This behavior indicates that a ground-state polaronic sector is not simply carried over into a transient polaron-like information partition.
Rather, the emergence of a phonon-dominated transient response involves a redistribution of correlations among excitonic, phononic, and photonic degrees of freedom.

The exciton-centered normalized QMI thus identifies a broad
polariton--polaron-like crossover in the transient information structure.
Within this broader crossover, a sharper boundary-adjacent structure appears
when the phonon subsystem is used as a reference.
We discuss this phonon-centered redistribution structure in the next
subsection.

\subsection{Phonon-centered information redistribution ridge}
\label{sec:phononridge}

The exciton-centered partition discussed in the previous subsection reveals a broad polariton--polaron-like crossover in the transient correlation structure.
We now use the phonon subsystem as the reference and examine how the phonon information is shared between the exciton and photon subsystems.
This change of reference is useful since the phonon and photon subsystems are not directly coupled in the Hamiltonian; any phonon--photon correlation must be generated through the excitonic degrees of freedom.
As shown below, the phonon-centered analysis reveals a sharper boundary-adjacent structure that is not visible in the photon-absorption map.
This structure should therefore not be interpreted as an enhancement of optical absorption, but as a redistribution feature that appears when the phonon information is resolved into exciton-side and photon-side correlations.

For the present purpose, we define the phonon-centered information partitions
\begin{equation}
\eta_{pt|pn}
=
\frac{\overline{I}_{pn,pt}}{2\overline{S}_{pn}},
\qquad
\eta_{e|pn}
=
\frac{\overline{I}_{e,pn}}{2\overline{S}_{pn}},
\label{eq:eta_phonon}
\end{equation}
which satisfy $\eta_{pt|pn}+\eta_{e|pn}=1$.
In addition, we introduce the redistribution ratio
\begin{equation}
R_{pn}
=
\frac{\overline{I}_{pn,pt}}{\overline{I}_{e,pn}}
=
\frac{\eta_{pt|pn}}{\eta_{e|pn}} .
\label{eq:Rn}
\end{equation}
While $\eta_{pt|pn}$ measures the fraction of phonon information shared with the photon subsystem, $R_{pn}$ compares the photon-side and exciton-side allocations of phonon information.

\begin{figure}
\scalebox{0.24}{\includegraphics*{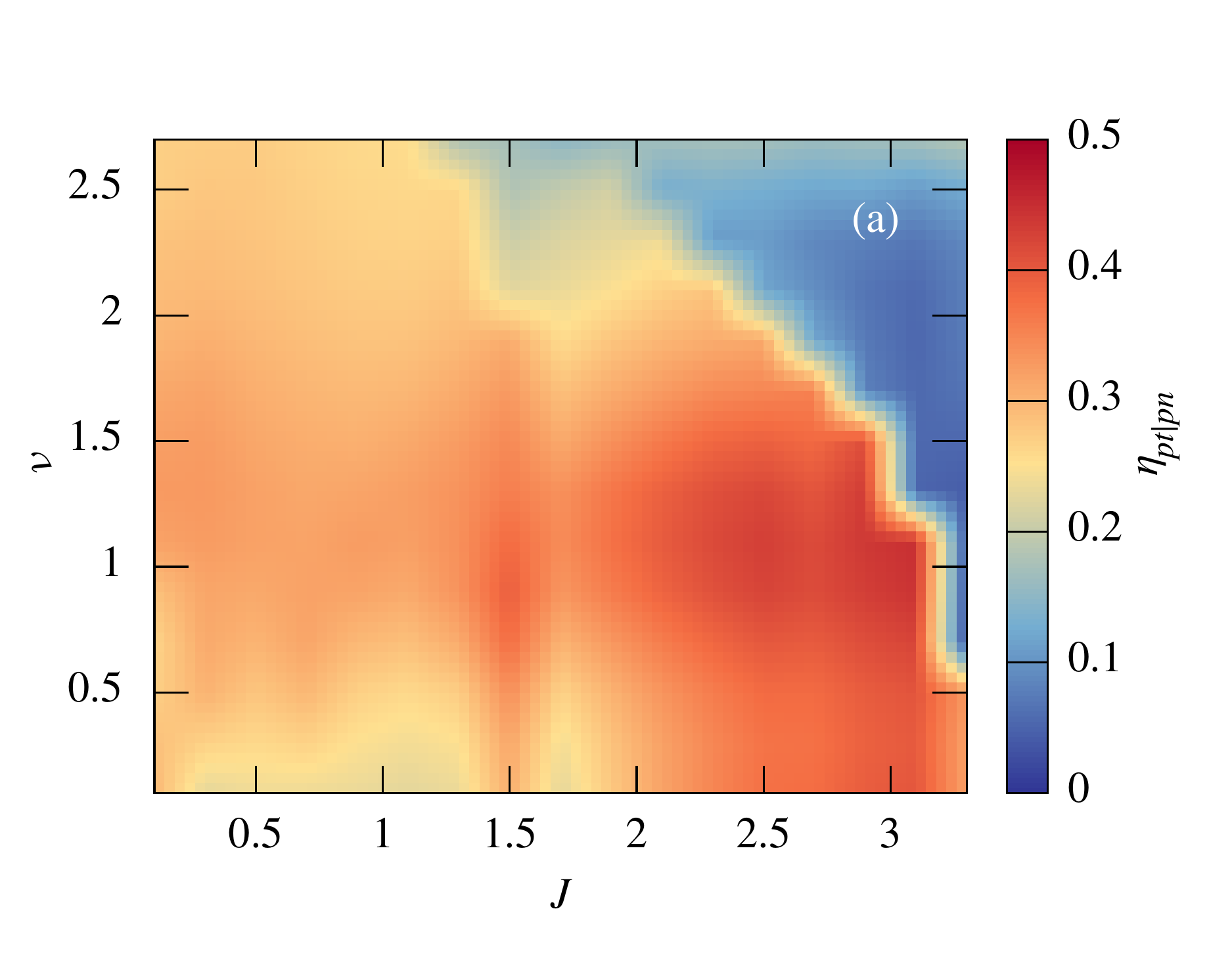}}
\scalebox{0.24}{\includegraphics*{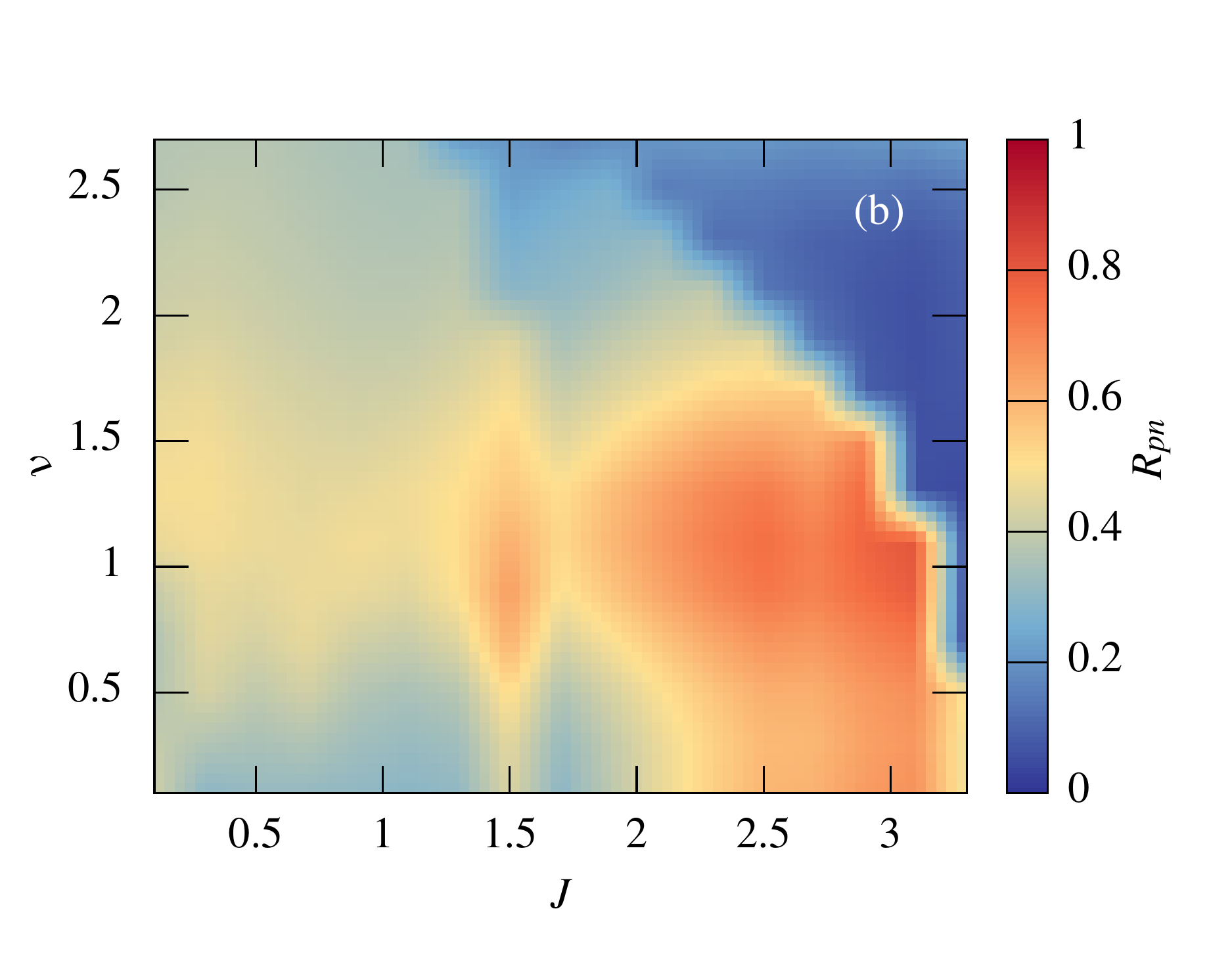}}
\scalebox{0.7}{\includegraphics*{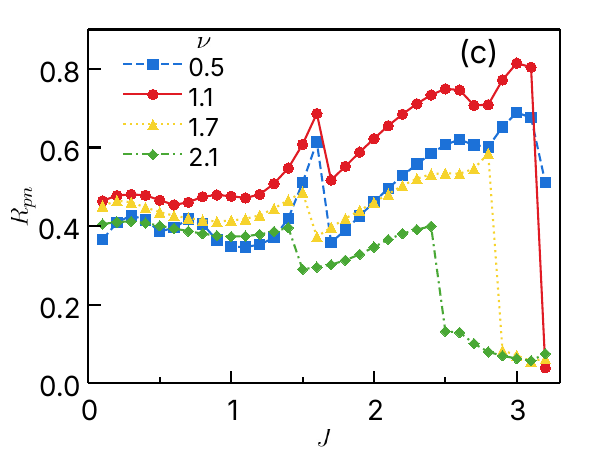}}
\caption{Phonon-centered information redistribution.
(a) $\eta_{pt|pn}$ and (b) $R_{pn}$ in the $J$-$\nu$ plane.
(c) Line cuts of $R_{pn}$ for $\nu = 0.5$, 1.1, 1.7, and 2.1.
The ridge appears near the I/II boundary on the region-I side.}
\label{fig:phonon_routing}
\end{figure}
Figure~\ref{fig:phonon_routing}(a) shows the map of $\eta_{pt|pn}$ in the $J$-$\nu$ plane.
A ridge-like enhancement appears near the low-$\nu$ part of the I/II boundary, slightly on the region-I side.
Although $\overline{S}_{pn}$ itself is not large in this region, a relatively large fraction of the available phonon information is shared with the photon subsystem.
Thus, the ridge reflects a preferential allocation of a limited phonon-information component to the photon side.
This feature is absent from the photon-absorption map shown in Fig.~\ref{fig:qmi_maps}(d), indicating that the ridge is not an optical-activity feature.

The same structure is sharpened in the redistribution ratio $R_{pn}$, shown in Fig.~\ref{fig:phonon_routing}(b).
As in Fig.~\ref{fig:phonon_routing}(a), the ridge in $R_{pn}$ runs approximately along the I/II boundary and does not coincide with it.
The displacement of the ridge toward region I is important.
On the low-$\nu$ side of region I, the photon channel is accessible whereas the phonon channel is only weakly involved.
Thus, little phonon information is available to be shared with photons.
In region II, the static polaronic character enhances exciton--phonon dressing, and the available phonon information is mainly shared with the exciton subsystem.
Near the I/II boundary on the region-I side, the photon channel remains accessible while the proximity to the polaronic sector makes the phonon channel available before the phonon information is predominantly allocated to exciton--phonon dressing.
The ridge therefore marks a boundary-adjacent region where a limited amount of phonon information is preferentially shared with the photon subsystem.
This interpretation does not require a large phonon entropy in the ridge region. 
Rather, the relevant point is the onset of a local exciton--phonon correlation channel near the I/II boundary.

This interpretation also explains why the ridge is not accompanied by a corresponding enhancement in the absorbed photon number.
The absorption map characterizes how strongly the photon mode participates in the dynamics, whereas $\eta_{pt|pn}$ and $R_{pn}$ characterize how the phonon information generated during the coherent evolution is distributed.
Thus, similar levels of photonic activity can lead to different transient information structures.
The ridge reflects the routing of weak phonon information toward the photon channel, rather than an increase in the total optical excitation.

Line cuts of $R_{pn}$, shown in Fig.~\ref{fig:phonon_routing}(c), confirm that the ridge is not an artifact of the color-map interpolation.
For representative values of $\nu$, $R_{pn}$ exhibits a local maximum near the I/II boundary on the region-I side.
As $\nu$ increases further, this maximum is suppressed, consistent with the transfer of phonon information toward exciton--phonon dressing.
The line cuts therefore support the interpretation of the ridge as a boundary-adjacent redistribution structure rather than as a separate static sector boundary.

The physical origin of the ridge can be related to a finite-energy polaronic excitation. 
Since the ridge appears close to, but not exactly on, the I/II boundary, the minimal variational polaron ansatz in Eq.~(\ref{var}) provides a good reference to a low-lying single-exciton excitation inside region I.
In addition, the ridge is closely related to the phonon-mediated information redistribution, which shows that phonon excitation plays an essential role in its formation.
We therefore examine whether Eq.\ (\ref{var}) satisfies a one-phonon resonance condition, which is expressed as 
\begin{equation}
4J+\omega =
-4J e^{-\alpha^2}
+\varepsilon
+2\nu\alpha
+\omega\alpha^2 ,
\label{varexc}
\end{equation}
where $\alpha$ is optimized variationally. 
This equation replaces the ground-state boundary condition in Eq.~(\ref{bnd}) by a one-phonon excitation
condition. 
For $\nu=0.9$, we obtain $J\simeq1.55$ and $\alpha\simeq-0.13$, in good agreement with the ridge position
$J\simeq1.5$ within the resolution of the numerical mesh. This agreement
suggests that the ridge is assisted by a region-II-like polaronic excitation
whose energy is resonant with the phonon energy.

To examine whether this variational configuration remains a well-defined
excitation, we calculate the strength function
\begin{equation}
A_{\rm II}(E) = \sum_n
|\langle \psi_n|\Phi_{\rm II}\rangle|^2
\delta[E-(E_n-E_a)] ,
\label{strengthfunc}
\end{equation}
where $|\Phi_{\rm II}\rangle$ is the optimized region-II variational state,
and $|\psi_n\rangle$ are the exact eigenstates in the corresponding
matter-sector Hilbert space. 
The energy is measured from the region-I ground-state energy $E_a=4J$. 
We note that this quantity is not an optical absorption spectrum, and that it measures the spectral distribution of the region-II-like configuration over the exact matter eigenstates.
The numerical details of the overlap strength-function analysis are given in the Supplemental Material \cite{supp}, where we also discuss its relation to the spectral-weight representation used in Lanczos calculations of dynamical correlation functions \cite{GaglianoBalseiro1987PRL}.

\begin{figure}
\scalebox{0.7}{\includegraphics*{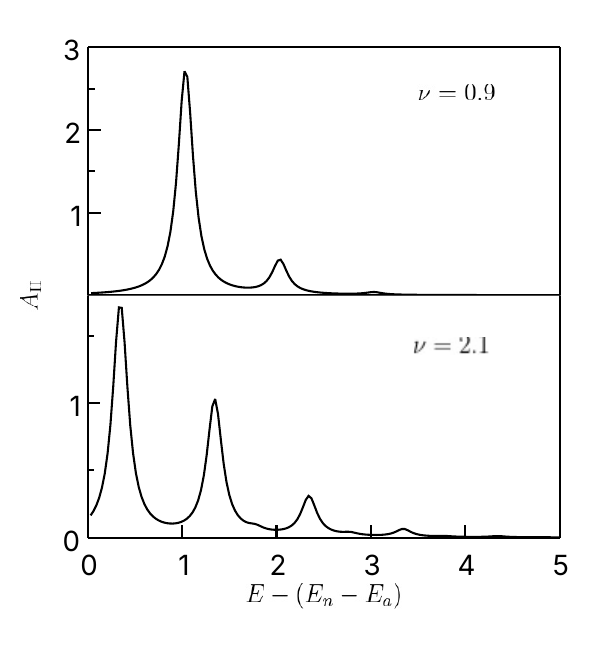}}
\caption{Strength function $A_{\rm II}(E)$ of the region-II variational state for $\nu=$0.9 and 2.1. 
The delta function is broadened by a Lorentzian of width 0.1. 
At $\nu=$0.9, the dominant peak appears at $E \simeq \omega$, whereas at $\nu=2.1$ the spectral weight shifts to lower energy and develops a ladder-like vibronic structure.}
\label{fig:strength}
\end{figure}
Figure~\ref{fig:strength} compares $A_{\rm II}(E)$ for $\nu=0.9$, where the ridge is most pronounced, and $\nu=2.1$, where the ridge has disappeared.
The delta function is broadened by a Lorentzian of width $0.1$. 
For $\nu=0.9$, the dominant peak appears at $E\simeq\omega$, indicating an isolated one-phonon-like excitation. In contrast, for $\nu=2.1$, the dominant spectral weight shifts to lower energy and additional peaks appear with approximately $\omega$-spaced intervals. 
Thus the disappearance of the ridge is not due to the suppression of phonon-assisted dynamics. 
Rather, the isolated one-phonon resonance is replaced by a broader vibronic manifold, and the phonon channel loses its selectivity.

These results indicate that the ridge is rooted in a finite-energy spectral structure of the electron--phonon subsystem. Its role, however, becomes visible only in the photon-coupled transient dynamics, where QMI resolves the redistribution of correlations among the electronic, phononic, and photonic degrees of freedom. 
In this sense, the ridge is a hidden dynamical correlation structure rather than a distinct ground-state phase.

The phonon-centered analysis therefore adds a new layer to the transient correlation map. 
The photon-absorption map gives the conventional optical background, and the exciton-centered partition identifies a broad polariton--polaron-like crossover. 
By contrast, the phonon-centered partition resolves a sharper redistribution ridge that is hidden from both the static ground-state classification and the absorption map. 
This hierarchy demonstrates that light irradiation redistributes correlations not only within the material sector but also between the material degrees of freedom and the photon mode, underscoring the need to treat the excitonic, phononic, and photonic sectors as a coherent quantum system.

\section{Discussion}

\subsection{Correlation-based framework for coherent optical-material dynamics}

The present results show that QMI-based information partition provides a correlation-based framework for characterizing coherent optical-material dynamics.
Conventional optical observables, such as the absorbed photon number, quantify how strongly the photon mode participates in the dynamics.
By contrast, the QMI-based quantities characterize how the correlations generated by light irradiation are shared among the excitonic, phononic, and photonic subsystems.
The transient state is therefore classified not only by the amount of optical excitation, but also by the partition of correlations among the relevant quantum degrees of freedom.

This viewpoint is particularly useful when polariton-like and polaron-like correlations coexist.
The exciton-centered partition gives a broad polariton--polaron-like classification of the transient response by identifying whether excitonic information is mainly shared with the photon subsystem or with the phonon subsystem.
The phonon-centered partition provides a complementary view of the same quantum state because it resolves how the phonon information is allocated between exciton-side dressing and photon-side correlation.
The redistribution ridge discussed in Sec.~\ref{sec:phononridge} is found in this phonon-centered decomposition and therefore represents a structure that is not resolved by the exciton-centered partition alone.

The ridge is not merely a feature missing from the photon-absorption map.
It indicates a change in the physical role of the photon subsystem.
Near the ridge, a limited amount of phonon-related information is preferentially shared with the photon subsystem, even though the total absorbed photon number is not enhanced.
In this sense, the ridge marks a parameter regime in which the photon sector carries an enhanced share of lattice-related information generated during the coherent transient evolution.

These observations clarify the role of the correlation-based framework introduced here.
Conventional optical descriptions often regard light as a probe or drive and matter as the target of the response.
This separation is appropriate when the optical field mainly probes or excites a material response.
In the coherent regime considered here, however, the photon mode itself participates as a quantum subsystem of the transient state.
As a result, the relevant physics is not exhausted by the material response to light, and it also includes the partition of correlations between optical and material degrees of freedom.
This perspective is compatible with the broader idea of on-demand quantum materials, where light is used not only to probe a preexisting material state but also to create and control a transient quantum state.
The QMI-based information partition is therefore particularly suited to characterizing such jointly formed light--matter states, for which the conventional probe--target separation is no longer sufficient.

\subsection{Spectral origin of the redistribution ridge}

The variational and strength-function analyses in Sec.~\ref{sec:phononridge} provide a microscopic interpretation for the redistribution ridge. 
The ridge appears when a region-II-like finite-energy polaronic excitation has dominant spectral weight near the one-phonon energy.
This connects the QMI-resolved redistribution structure to the excitation spectrum of the electron--phonon subsystem.

At the same time, the ridge should not be identified simply with a matter eigenstate.
The strength function $A_{\rm II}(E)$ shows that the region-II-like configuration remains a relevant spectral component, whereas the ridge itself appears in the photon-coupled transient dynamics of the full electron--phonon--photon system. 
The finite-energy excitation therefore acts as a microscopic anchor for the ridge, while QMI reveals its role as a transient correlation pathway involving the photon subsystem.

For smaller $\nu$, the spectral weight of the region-II-like variational state is concentrated near the one-phonon energy, and the corresponding phonon-assisted channel is selective.
Consequently, this selectivity produces the enhancement in $\eta_{pt|pn}$ and $R_{pn}$. 
For larger $\nu$, on the other hand, the strength function develops a vibronic ladder, and the spectral weight is distributed over several eigenstates. 
Although the phonon channel remains active, it loses its selectivity, and the ridge disappears even though electron--phonon correlations themselves are not suppressed as a result.

This analysis also clarifies the meaning of the term hidden in the present work. 
The ridge is not a distinct symmetry-broken phase, nor is it invisible in principle to spectral analysis. 
Rather, it is hidden from the ground-state classification and from population-based optical observables. 
It becomes apparent when the transient state is analyzed through the QMI-resolved partition of correlations, and, in this sense, the ridge is a hidden dynamical correlation structure rather than a distinct ground-state phase.

\subsection{Finite-size aspects and minimal mechanism of correlation redistribution}

The present calculation is performed for a four-site system and should not be interpreted as a finite-size extrapolation of a thermodynamic phase transition. 
The purpose of the four-site model is to provide a minimal fully quantum setting in which the excitonic, phononic, and photonic correlation channels can be treated on the same footing. 
Within this setting, the ground-state sector structure, the optical response, and the transient information partition can be compared without semiclassical or mean-field approximations.

The I/II boundary has a relatively robust local origin, i.e., it is governed by the energetic competition between excitation creation, exciton delocalization, and phonon-induced dressing. 
The minimal variational polaron ansatz captures this balance and provides a useful reference for the onset of polaronic correlations. 
The redistribution ridge appears near this boundary because a region-II-like polaronic configuration becomes relevant as a finite-energy excitation inside region I.

Larger systems will introduce additional structures beyond the present minimal model. 
Multiple-polaron configurations, spatial correlations among polarons, domain-like structures, and wave-vector selectivity of phonon modes will modify the detailed shape and sharpness of the ridge. 
These effects are expected to be more sensitive to system size than the elementary correlation redistribution mechanism discussed in this study. 
The present results should therefore be read as identifying a minimal mechanism of transient correlation redistribution associated with the onset of polaronic correlations, rather than as establishing the full thermodynamic structure of a photoinduced phase.

We stress that this finite-size limitation does not diminish the role of the correlation-based framework. 
QMI does not remove finite-size effects; instead, it formulates the question in terms of how correlations are shared among physically chosen subsystems. 
The four-site system is therefore used here not as a miniature bulk material, but as a minimal coherent quantum system in which exciton, phonon, and photon correlation channels can be explicitly compared.

\subsection{Implications for observable properties}

The present study does not claim that QMI itself is directly measured in standard optical experiments. 
Rather, QMI identifies parameter regimes in which the character of the transient state changes even when conventional optical measures, such as the absorbed photon number, remain similar.
It therefore provides guidance for constructing observable projections of the underlying correlation structure.

Coherent phonon experiments show how lattice dynamics can be projected onto optical observables.
In particular, coherent phonons generated by femtosecond light pulses have long been used to project lattice dynamics onto time-resolved optical signals \cite{Merlin1997SSC}.
Time-dependent coherent-phonon spectra can further serve as all-optical probes of changes in the lattice potential and symmetry during photoinduced structural dynamics \cite{Wall2012NatCommun}. 
Conversely, mode-selective vibrational excitation demonstrates that selected lattice coordinates can actively control electronic phases \cite{Rini2007Nature}, and frequency-domain time-resolved photoemission has shown that electron--phonon couplings during a photoinduced transition can be resolved in a mode- and band-selective manner \cite{Suzuki2021PRB}.
In the present context, these experiments motivate the search for observable projections that are sensitive not only to photon population or lattice displacement, but also to photon--phonon and photon--matter correlation channels.

The redistribution ridge suggests that phonon-related information is transiently encoded in the photon sector.
This points to observables that are sensitive not only to photon population, but also to photon correlations and photon--matter correlations.
Possible examples include photon-number fluctuations, photon--phonon connected correlations, phase-sensitive optical signals, and emission properties that depend on the lattice component of the transient state. 
In particular, if the ridge corresponds to enhanced phonon information in the photon sector, then photon observables sensitive to phonon-induced phase or fluctuation channels may change even without a corresponding enhancement in the total absorbed photon number.

Thus, coherent light irradiation should be viewed not only as population transfer, but also as a process that creates a transient partition of correlations among material and optical degrees of freedom, particularly when quantum coherence between material and photons remains.
The redistribution ridge demonstrates this point: it marks a resonant condition under which lattice-related information is selectively encoded in the photon sector.
This is the type of structure that is difficult to identify from absorption or static classification alone, and becomes visible when the coherent transient state is analyzed through a correlation-based information partition.

\section{Conclusions}

We have studied transient correlation redistribution in a closed exciton--phonon--photon system.
The model treats excitonic, phononic, and photonic degrees of freedom as coherent quantum subsystems and therefore provides a minimal setting for comparing the ground-state sector structure with the information structure generated after photon excitation.

By analyzing subsystem-resolved quantum mutual information (QMI), we showed that the photoexcited state cannot be characterized only by the absorbed photon number or by the ground-state classification.
The absorbed photon number provides a conventional measure of photonic activity, whereas QMI resolves how the correlations generated by light irradiation are shared among the exciton, phonon, and photon subsystems.
The exciton-centered information partition identifies a broad crossover between polariton-like and polaron-like transient responses, depending on whether excitonic information is mainly shared with the photon or phonon subsystem.

Using the phonon subsystem as the reference, we further found a boundary-adjacent redistribution ridge near the I/II boundary on the region-I side.
This ridge is absent from both the ground-state sector map and the photon-absorption map, and therefore represents a hidden transient correlation structure rather than a static sector boundary or an enhancement of optical absorption.
Physically, the ridge marks a regime in which a limited amount of phonon-related information is preferentially shared with the photon subsystem before it is predominantly allocated to exciton--phonon dressing.
In this sense, the ridge indicates that lattice-related information can be transiently encoded in the photon sector without a corresponding enhancement of the total absorbed photon number.

A variational strength-function analysis connects this redistribution ridge to a region-II-like finite-energy polaronic excitation.
The ridge is most pronounced when the dominant spectral weight of this excitation lies near the one-phonon energy, whereas it disappears when the spectral weight is distributed over a vibronic ladder.
This result shows that the ridge is rooted in the spectral structure of the electron--phonon subsystem, while its role as a correlation-transfer pathway becomes visible only in the photon-coupled transient dynamics.

These results demonstrate that QMI-based information partition provides a correlation-based framework for characterizing coherent light-induced transient states.
Unlike population-based observables or selected microscopic correlation functions, this framework compares different correlation channels on the same footing and classifies the transient state by how optically generated correlations are partitioned among the relevant quantum degrees of freedom.
It is therefore particularly useful when optical and material degrees of freedom jointly form a coherent transient state.

Although the present calculation was performed for a four-site system, the approach is not limited to this specific model.
The same framework can be applied to a broad class of electron--lattice systems and, more generally, to coherent many-body systems composed of multiple interacting subsystems.
Future work should clarify how the redistribution structures identified here evolve in larger systems and how they are projected onto experimentally accessible quantities such as photon statistics, phonon-sensitive optical signals, and connected correlation functions.

\acknowledgments
The author is grateful to A. Kikuchi and N. Aoyagi for valuable discussions.
Numerical calculations were performed on the facilities at the Supercomputer Center, Institute for Solid State Physics, the University of Tokyo, Japan.
This work is partly supported by JSPS KAKENHI Grant Numbers 25K08502, 26H02255, and 26K01417.

\bibliographystyle{apsrev4-2}
\bibliography{references_prr_updated}

\end{document}